\documentclass[a4paper,11pt,oneside]{article}

\usepackage[english]{babel}
\usepackage{amsmath, amssymb}
\usepackage{enumerate}
\usepackage{enumitem}

\begin{document}

\setcounter{secnumdepth}{3}

\title{Corrected XVA Modelling Framework \\ and Formulae for KVA and MVA\footnote{The views expressed in this article are those of the author only, and do not necessarily represent the opinions of Lloyds Banking Group.}}
\author{Antti K. Vauhkonen\footnote{Contact: antti.vauhkonen@lloydsbanking.com}}
\date{This version was first posted on SSRN on July 25, 2018.}
\maketitle 

\begin{abstract}
We discuss and clarify the XVA modelling framework specified in the paper ``MVA by replication and regression" (\textit{Risk}, May 2015) for including  bilateral credit risk and funding costs in derivative pricing, and in doing so we rectify two key errors in the valuation adjustments accounting for costs of capital and initial margin, and present corrected formulae for KVA and MVA.  
\end{abstract}


\section{XVA modelling by semi-replication strategy}

\par In this article we examine the semi-replication framework introduced by Burgard and Kjaer in \cite{BK13} for incorporating default risks of counterparties and costs of funding hedge positions in the valuation of derivative contracts, subsequently extended by Green \textit{et al.} in \cite{GKD14} and \cite{GK15} to take into account costs of capital and funding initial margin posted under a CSD agreement. 

\par Adopting the same notation as in \cite{GK15}, and suppressing in most instances the explicit time dependence of all the variables, we consider the value to the bank of a derivative contract (or a portfolio of derivative contracts) $\hat{V}$ that depends on the price of an underlying asset $S$ -- which we may think of as a stock paying continuous cash dividends at rate $\gamma_{S}$ -- and is also dependent on the default statuses of the bank and its counterparty, i.e. 
\begin{equation}
\hat{V}=\hat{V}(t,S,J_{B},J_{C})
\end{equation}
where $J_{B}$ and $J_{C}$ are default indicators for the bank and the counterparty, respectively. The underlying asset is assumed to follow a geometric Brownian  motion: 
\begin{equation}
dS=\mu Sdt+\sigma SdW \,,
\end{equation}
and the default indicators are modelled using Poisson processes.

\par For hedging bilateral credit risk, we shall use two zero-coupon bonds of different seniorities issued by the bank (with the first bond assumed to be a subordinated one), and a zero-coupon bond issued by the counterparty whose prices follow the stochastic processes given below:              
\begin{subequations}
\begin{align}
dP_{i} \, & =r_{i}P_{i}dt-(1-R_{i})P_{i}dJ_{B} \\
dP_{C} & =r_{C}P_{C}dt-P_{C}dJ_{C} 
\end{align}
\end{subequations} 
where $r_{i}$ and $R_{i}$ for $i\in\{1,2\}$ are the yields and recovery rates on the bank bonds, and $r_{C}$ is the yield on the counterparty bond. Note from equation (3b) that the recovery rate on the counterparty bond is assumed to be zero, i.e. the counterparty bond is assumed to be a subordinated one. Thus, the bond yields can be expressed as 
\begin{subequations}
\begin{align}
r_{i}\, & =r+(1-R_{i})\lambda_{B} \\
r_{C} & =r+\lambda_{C}
\end{align}
\end{subequations}
where $r$ is the risk-free rate, and $\lambda_{B}$ and $\lambda_{C}$ are the hazard rates for the bank and the counterparty, respectively. 

\par Using these instruments a portfolio is constructed to hedge the derivative position consisting of $\delta$ units of the underlying asset, $\alpha_{C}$ counterparty bonds, $\alpha_{1}$ subordinated bank bonds, $\alpha_{2}$ senior bank bonds, and a cash amount $\beta$. Thus, denoting the value of the hedging portfolio by $\Pi$, we have 
\begin{equation}
\Pi = \delta S + \alpha_{C}P_{C} + \alpha_{1}P_{1} + \alpha_{2}P_{2} + \beta \,.
\end{equation}

\par We assume that the hedge positions in stock and counterparty bonds are funded through repo transactions, and also that the following funding condition holds at all times: 
\begin {equation}
\hat{V}+\alpha_{1}P_{1}+\alpha_{2}P_{2}+I_{B}-X-\phi K = 0 
\end{equation} 
where $I_{B}$ is the initial margin posted by the bank, $X$ is the variation margin collateral held by the bank, $K$ is the regulatory capital requirement for holding the derivative position, and $\phi \in [0,1]$ is the fraction of the raised capital that has been made available for funding purposes. This means that any shortfall between the funding required for the derivative, initial margin and a long position in bank bonds of one seniority (in accordance with the hedging strategy) and the funding provided by received collateral and capital is financed by the bank issuing bonds of the other seniority. 
\par Therefore, we can write $\beta = \beta_{S} + \beta_{C} + \beta_{X} + \beta_{K} + \beta_{I_{B}}$ where $\beta_{S}=-\delta S$, $\beta_{C}=-\alpha_{C}P_{C}$, $\beta_{X}=-X$, $\beta_{K}=-\phi K$ and $\beta_{I_{B}}=I_{B}$ with the following growth rates:     
\begin{subequations}
\begin{align}
d\beta_{S} & = (\gamma_{S}-q_{S})\delta Sdt \\
d\beta_{C} & = -q_{C}\alpha_{C}P_{C}dt \\
d\beta_{X} & = -r_{X}Xdt \\
d\beta_{K} & = -\gamma_{K}\phi Kdt \\
d\beta_{I_{B}} & = r_{I_{B}}I_{B}dt 
\end{align}
\end{subequations}
where $q_{S}$ and $q_{C}$ are the repo rates for the stock and counterparty bonds, respectively, $r_{X}$ is the rate of interest payable on variation margin collateral, $\gamma_{K}$ is the cost of capital, and $r_{I_{B}}$ is the rate of interest earned on posted initial margin.

\
\\ 
\textbf{Remarks:} 

\begin{enumerate}[label=\arabic*.]  
\item The repo rates $q_{S}$ and $q_{C}$ in equations (7a) and (7b), respectively, are assumed to be adjusted for any applicable haircuts, i.e. for the fact that in a repo transaction the collateral asset has to be partially funded at a higher, unsecured rate due to the haircut applied to its value. 
\item In equation (7b), there is no term for cash return on counterparty bonds, since they are assumed to be zero-coupon bonds. Moreover, it is implicitly assumed that both bank and counterparty bonds mature after the maturity date of the derivative contract.  
\item By the definition of $\phi$, the cash account $\beta_{K}$ associated with capital funding is $-\phi K$, not $-K$. Thus, $\phi$ is missing from the right hand side of equation (11) in \cite{GK15} that corresponds to our equation (7d) above. Even though the capital requirement $K$ for the derivative $\hat{V}$ incurs a running expense of $\gamma_{K}K$ per unit time to the trading desk holding the position (or is borne centrally at a higher business unit or divisional level), only the costs of funding the derivative, initial and variation margin collateral and the hedge positions in the underlying asset, bank and counterparty bonds, as given by the growth rates of the cash accounts in equations (7a)--(7e), are factored into this pricing analysis for $\hat{V}$. 
\item In equation (7e), there is no term for the initial margin posted by the counterparty, $I_{C}$, since under regulatory rules initial margin collateral must be held in a segregated account, and cannot be reused by the receiving counterparty for funding purposes. Moreover, from a legal point of view, there is a fundamental difference between variation margin collateral and initial margin collateral: for, posting variation margin involves legal transfer of title to the collateral, which enables the receiving counterparty to reuse it freely, and, analogously to a loan, interest is payable by the receiving counterparty on variation margin collateral, as reflected in equation (7c); whereas initial margin collateral is only pledged to the receiving counterparty with the posting counterparty remaining the legal owner of the collateral, and any cash returns on initial margin collateral while in third-party custody -- any interest accrued on cash funds posted as initial margin collateral and placed in a general deposit account of the custodian bank, or any coupon payments on debt securities or dividends on stocks held in segregated accounts at third-party custodians -- are payable by either the custodian bank (in the case of cash collateral) or the issuers of the securities (in the case of non-cash collateral), not the counterparty receiving (security interest in) the initial margin collateral. Consequently, there is no cash account $\beta_{I_{C}}$ associated with initial margin \textit{received} by the bank, and there is no $-r_{I_{C}}I_{C}dt$ term on the right hand side of equation (7e), as has erroneously been included in the corresponding equation (12) in \cite{GK15}. 
\item The use of initial margin, both posted and received, in close-out netting is correctly reflected in equations (9a) and (9b) below. For example, $I_{C}$ does not appear in the expression for the close-out value $g_{B}$ in the event of bank default, as in this case any initial margin that had been posted by the counterparty and held in a segregated account by a third-party custodian would simply be returned to the counterparty and not netted against the derivative value. Similarly, $I_{B}$ is absent from the expression for $g_{C}$.  
\end{enumerate}

\par Now, applying It\^{o}'s lemma to $\hat{V}$, and assuming that the hedging portfolio is self-financing, we obtain  
\begin{equation}
\begin{split}
d\hat{V} + d\Pi = \Bigg ( & \frac{\partial \hat{V}}{\partial t} + \tfrac{1}{2}\sigma^2 S^2\frac{\partial^2\hat{V}}{\partial S^2} + (\gamma_{S}-q_{S})\delta S + (r_{C} - q_{C})\alpha_{C}P_{C} \\
 & + r_{1}\alpha_{1}P_{1}  + r_{2}\alpha_{2}P_{2} - r_{X}X - \gamma_{K}\phi K +r_{I_{B}}I_{B} \Bigg ) dt \\
 & + \Bigg ( \frac{\partial \hat{V}}{\partial S} + \delta \Bigg )dS + \epsilon_{h}dJ_{B} + \big(g_{C} -\hat{V} - \alpha_{C}P_{C}\big)dJ_{C} 
\end{split}
\end{equation}
where $\epsilon_{h} = g_{B} + R_{1}\alpha_{1}P_{1} + R_{2}\alpha_{2}P_{2} - X - \phi K + I_{B}$ and 
\begin{subequations}
\begin{align}
g_{B} & = (V-X+I_{B})^{+} + R_{B}(V-X+I_{B})^{-} + X - I_{B} \\
g_{C} & = R_{C}(V-X-I_{C})^{+} + (V-X-I_{C})^{-} + X + I_{C}
\end{align}
\end{subequations}
are the close-out values to the bank of the collateralised derivative position in the event of the bank or the counterparty defaulting, $R_{B}$ and $R_{C}$ are the recovery rates on the collateralised derivative position upon bank and counterparty default, respectively, and $V$ is the risk-free value of the derivative. As in \cite{GK15}, we shall assume from now on that $q_{C}$ is equal to the risk-free rate $r$.

\par Thus, by setting $\delta = -\frac{\partial \hat{V}}{\partial S}$ and $\alpha_{C} =(g_{C} - \hat{V})/P_{C}$ in equation (8) above eliminates risk to market movements in the value of the underlying asset and default risk of the counterparty, and $d\hat{V} + d\Pi = 0$ whenever $dJ_{B}=0$, i.e. as long as the bank has not defaulted, provided that the valuation adjustment $U := \hat{V} - V$ satisfies the following partial differential equation: \begin{equation}
\begin{split}
& \frac{\partial U}{\partial t} + \tfrac{1}{2}\sigma^2 S^2\frac{\partial^2 U}{\partial S^2} + (q_{S}-\gamma_{S})S \frac{\partial U}{\partial S} - ( r +\lambda_{B} + \lambda_{C} ) U = \\
 & - \lambda_{B}(g_{B}-V) - \lambda_{C}(g_{C}-V) + \lambda_{B}\epsilon_{h} + s_{X}X  + (\gamma_{K} - r)\phi K - s_{I_{B}}I_{B} 
\end{split}
\end{equation}
where $s_{X}:=r_{X}-r$ and $s_{I_{B}}:=r_{I_{B}}-r$, subject to the terminal condition $U(T,S)=0$ for all values of S where $T$ is the maturity of the derivative contract.

\section{XVA formulae}

\par Applying the Feynman-Kac formula to equation (10), and, writing $U(t)$ as the sum of its component elements, gives 
\begin{equation}
U(t) = CVA(t) + DVA(t) + FCA(t) + ColVA(t) + KVA(t) + MVA(t)
\end{equation}
with 
\begin{equation*}
\begin{split}
CVA(t) & = - (1-R_{C})\int_{t}^{T} D(t,u) \lambda_{C}(u)\mathbb{E}_{t}\left[(V(u)-X(u)-I_{C}(u))^{+}\right]du  \\
DVA(t) &  = - (1-R_{B})\int_{t}^{T} D(t,u) \lambda_{B}(u)\mathbb{E}_{t}\left[(V(u)-X(u)+I_{B}(u))^{-}\right]du  \\ 
FCA(t) & = - \int_{t}^{T} D(t,u) \lambda_{B}(u)\mathbb{E}_{t}\left[g_{B}(u)+R_{1}\alpha_{1}(u)P_{1}(u)+R_{2}\alpha_{2}(u)P_{2}(u)\right]du  \\
ColVA(t) & = - \int_{t}^{T} D(t,u) \{ r_{X}(u) - (r(u) + \lambda_{B}(u)) \} \mathbb{E}_{t}\left[X(u)\right]du  
 \\
KVA(t) & = - \int_{t}^{T} D(t,u) \{ \gamma_{K}(u) - (r(u) + \lambda_{B}(u)) \} \mathbb{E}_{t} \left[ \phi K(u) \right] du 
 \\
MVA(t) & = + \int_{t}^{T} D(t,u) \{ r_{I_{B}}(u) - (r(u) + \lambda_{B}(u)) \} \mathbb{E}_{t} \left[ I_{B}(u) \right] du
\end{split}
\end{equation*}
where $D(t,u) := e^{-\int_{t}^{u}(r(v)+\lambda_{B}(v)+\lambda_{C}(v))dv}$.

\
\par It should be noted that the above expressions for KVA and MVA differ from those given in \cite{GK15}, and take the form that one would intuitively expect: namely, KVA and MVA are calculated, respectively, as the spread between the cost of capital (return on posted initial margin) and the yield on a zero-recovery subordinated bank bond integrated over the expected capital funding (posted initial margin) profile. Also, ColVA has the same form.



\end{document}